\documentclass[twocolumn,showpacs,amsfonts,aps,prc,nofootinbib,floatfix,%
superscriptaddress]{revtex4}

\usepackage{float}

\usepackage{amsmath}
\usepackage{bm}
\usepackage{graphicx}

\usepackage{epsfig}
\newcommand{\beq}{\begin{equation}}
\newcommand{\eeq}{\end{equation}}
\newcommand{\bea}{\vspace{0.25cm}\begin{eqnarray}}
\newcommand{\eea}{\end{eqnarray}}

\def\lsim{\mathrel{\rlap{\lower4pt\hbox{\hskip1pt$\sim$}}
    \raise1pt\hbox{$<$}}}         
\def\gsim{\mathrel{\rlap{\lower4pt\hbox{\hskip1pt$\sim$}}
    \raise1pt\hbox{$>$}}}         

\begin{document}


\title{\Large\bf
Medium effects for hadron-tagged jets in 
proton-proton collisions
}

\date{\today}

\author{B.G. Zakharov}

\address{
L.D.~Landau Institute for Theoretical Physics,
        GSP-1, 117940,\\ Kosygina Str. 2, 117334 Moscow, Russia
}

\begin{abstract}
  We study the medium modification factor $I_{pp}$
within the light-cone path integral approach to induced gluon emission.
We use parametrization of the running coupling $\alpha_s(Q,T)$
which has a plateau around $Q\sim \kappa T$.
We calculate $I_{pp}$ with no free parameters
  using $\kappa$ fitted to the LHC data on the nuclear
  modification factor $R_{AA}$.
We find that the theoretical multiplicity dependence of the
ratio $I_{pp}/\langle I_{pp}\rangle$
for $5.02$ TeV $pp$ collisions
agrees reasonably with the recent preliminary data
from ALICE \cite{ALICE_Ipp}.

\end{abstract}
%

\maketitle
{\bf Introduction}.
The observations in $AA$ collisions at RHIC and the LHC
of the transverse flow effects and the strong suppression of high-$p_T$
hadron spectra (jet quenching) are the most compelling arguments in favor
of the quark-gluon plasma (QGP) formation in
$AA$ collisions (for reviews see, e.g., \cite{hydro2,hydro3,JQ_rev}).
The results of hydrodynamic analyses of the flow effects
support that the QGP is formed at the proper time $\tau_0\sim 0.5-1$ fm
\cite{hydro2,hydro3}.
Such a picture of the QGP formation
allows one to describe
the RHIC and LHC heavy ion data on the nuclear modification factor $R_{AA}$
(see, e.g., \cite{CUJET3,Armesto_LHC,Z_hl})  
in the pQCD jet quenching models with dominating contribution to the
jet modification from induced gluon
emission \cite{GW,BDMPS1,LCPI1,W1,GLV1,AMY1}.
The observation of the ridge effect
\cite{CMS_ridge,ATLAS_mbias} in $pp$ collisions at the LHC energies, suggests
that the QGP can be created in $pp$ collisions as well.
The mini QGP (mQGP) formation in $pp$ collisions 
is also supported by the steep growth  of
the midrapidity strange particle production
at charged multiplicity $dN_{ch}/d\eta\sim 5$ \cite{ALICE_strange}.
This agrees with the onset of the QGP regime at $dN_{ch}/d\eta \sim 6$, 
found in \cite{Camp1} from behavior of 
$\langle p_{T}\rangle$ as a function of multiplicity, 
employing Van Hove's arguments \cite{VH} that  
the phase transition should  lead to an anomalous multiplicity dependence of 
$\langle p_{T}\rangle$.
These estimates
of the critical multiplicity density for the onset of the mQGP formation regime
are smaller than the typical midrapidity charged multiplicity
of the soft (underlying-event (UE)) hadrons for jet events in
$pp$ collisions at the LHC energies --- $dN_{ch}/d\eta\sim 10-15$
(it is bigger than the ordinary minimum bias
multiplicity by a factor of $\sim 2-2.5$
\cite{Field}). 

{
  The typical size of the mQGP fireball created in $pp$ collisions
       should be $\sim 2-3$ fm. The Knudsen number, $Kn\sim \lambda/L$,  
for the mQGP may be estimated (as in \cite{Z_kn}) with the help
of the Drude formula and the lattice electrical conductivity
     of the QGP from lattice simulations \cite{sigma}. 
In this way for the mQGP in the minimum bias
$pp$ UEs at the proper time $\tau\sim 0.5(1)$ fm
one can obtain for quarks quite small values $Kn\sim 0.2(0.1)$.
For gluons the Knudsen number should be smaller by a factor of
$\sim C_A/C_F=9/4$. Such small values of $Kn$ support the collective
hydrodynamic behavior of the mQGP fireball in $pp$ collisions.
Note that it is possible (see e.g. \cite{RMK1512})
that the Knudsen number criterion
$Kn\ll 1$ is too crude and underestimates  the applicability of hydrodynamics. 
       In \cite{RMK1609} it was argued that the lower bound for the size of
       the mQGP with hydrodynamic behavior may be as small as $\sim 0.15$ fm.
There it was demonstrated that neither local near-equilibrium
       nor near-isotropy are required for successful description of
       $pp$ data  within viscous hydrodynamics. From the analysis
       of the data on $v_2$ it was concluded that in $pp$ collisions
       the hydrodynamics works at $dN_{ch}/d\eta\gsim 2$. This agrees
       well with the results of \cite{Spalinski}.
     }        
 
The mQGP formation in $pp$ collisions should lead
to some jet modification.
However, one can expect that the quenching effects in $pp$ collisions  should be
significantly smaller than in heavy ion collisions
due to lower temperature of the mQGP and due to strong
reduction of the induced gluon emission for a small size fireball.
The latter is closely related to the anomalously strong $L$-dependence 
of the radiative parton energy loss, $\Delta E_{r}$, in a finite-size QCD
matter \cite{BDMPS1} (as compared to predictions of the Bethe-Heitler
formula).
Fixed coupling calculations without the Coulomb effects
within the BDMPS approach \cite{BDMPS1}
give $\Delta E_{r}\propto L^2$ for a static QGP,
and $\Delta E_{r}\propto L$ \cite{BDMSexp} for an expanding QGP
with entropy density
$s\propto 1/\tau$ (as in the Bjorken model \cite{Bjorken} with purely
longitudinal expansion of the QGP). The linear $L$-dependence of $\Delta E_r$
for an expanding QGP remains approximately valid also for calculations  with accurate treatment of
the Coulomb effects with running $\alpha_s$ \cite{Z_pp13}.
Calculations of the medium modification factor $R_{pp}$
(which is not directly observable quantity)
within the
light-cone path integral (LCPI) approach \cite{LCPI1}
with accurate treatment of the Coulomb effects and running $\alpha_s$
give 
a small deviation of $R_{pp}$ from unity
at the LHC energies \cite{Z_hl} ($R_{pp}\sim 0.8$ at $p_T\sim 10$ GeV). 
For this reason observation of jet quenching in $pp$ collisions via a weak
modification of the $p_T$-dependence of hadron spectra is practically
impossible.
A promising observable for quenching effects in $pp$ collisions
is the variation with the UE activity of the medium modification
factor $I_{pp}$ for the photon-tagged jet fragmentation functions (FFs)
\cite{Z_pp_PRL}.
However, this measurement requires high
statistics due to a very small cross section. This problem is absent
for the modification factor $I_{pp}$ for the hadron-tagged jets.
The medium modification factor $I_{pp}$ for the di-hadron
production in $pp$ collisions
can be written similarly to $AA$ collisions
\cite{PHENIX_di-h}
\beq
I_{pp}(p_T^a,p_T^t,y^a,y^t)=\frac{Y^{pp}_m(p_T^a,p_T^t,y^a,y^t)}
{Y_v^{pp}(p_T^a,p_T^t,y^a,y^t)}\,,
\label{eq:10}
\eeq
where
$p_T^{a,t}$ and $y^{a,t}$ are the transverse momenta and rapidities
of the trigger ($h^t$) and the associated ($h^a$) hadrons,
$Y^{pp}_m$ is the per-trigger yield accounting the medium effects,
and $Y_v^{pp}$ is the per-trigger yield calculated ignoring the medium effects.
The per-trigger yields (similarly to $AA$
collisions \cite{Wang_di-h,PHENIX_di-h})
can be written in terms of the di-hadron (back-to-back) and one-hadron
inclusive cross sections as 
\beq
Y^{pp}_{m,v}(p_T^a,p_T^t,y^a,y^t)=\frac{d^4\sigma_{m,v}}{dp_T^adp_T^tdy^ady^t}
\Big/\frac{d^2\sigma_{m,v}}{dp_T^tdy^t}\,.
\label{eq:20}
\eeq
Of course, the denominator in (\ref{eq:10}) is unobservable.
But one can study
the UE multiplicity dependence of $I_{pp}$, say, by using the ratio of
the per-trigger yield to its minimum bias value.
Because we can reasonably expect that the UE multiplicity dependence of
the numerator and the denominator of (\ref{eq:20}) for $Y_v^{pp}$
is very similar,
and consequently $Y_v^{pp}/\langle Y_v^{pp}\rangle\approx 1$. Then we have
\beq
\frac{I_{pp}(p_T^a,p_T^t,y^a,y^t)}
     {\langle I_{pp}(p_T^a,p_T^t,y^a,y^t)\rangle}\approx
\frac{Y_m^{pp}(p_T^a,p_T^t,y^a,y^t)}
     {\langle Y_m^{pp}(p_T^a,p_T^t,y^a,y^t)\rangle}\,.
     \label{eq:30}
     \eeq
 This relation allows one to study the multiplicity dependence of $I_{pp}$
by measuring the per-trigger yield (which corresponds to the theoretical
$Y_m^{pp}$).     
Recently, this method has been used
by ALICE \cite{ALICE_Ipp}
in the first measurement of the variation
of $I_{pp}$ with the UE multiplicity  for the hadron-tagged
jets for $5.02$ TeV $pp$ collisions
(for the trigger hadron momentum
$8<p_T^t<15$ GeV, and the associated away side hadron
momentum $4<p_T^a<6$ GeV). It was found that $I_{pp}$ decreases monotonically
by about 15\% with increase of the UE activity in the range
$5\lsim dN_{ch}/d\eta\lsim 20$
(we use $dN_{ch}/d\eta$ for the whole  range of the azimuthal angle
$\phi$ and the transverse momentum which is bigger
by a factor of $\sim 4.4$ than
the transverse side charged multiplicity $N_{ch}^{TS}$
of \cite{ALICE_Ipp}
for the kinematic region $\pi/3\leq |\phi|\leq 2\pi/3$,
$|\eta|<0.8$, and $p_T>0.5$ GeV).
Such a decrease of $I_{pp}$ agrees qualitatively with the quenching effect
obtained in \cite{Z_pp_PRL}  for the jet energy $E=25$ GeV (which is of the
order of the jet energy for the ALICE trigger particle momentum region
\cite{ALICE_Ipp}).
For drawing a more definitive conclusion on whether the ALICE data
\cite{ALICE_Ipp} on $I_{pp}$ may be consistent with jet quenching in the mQGP,
it is of course highly desirable to perform calculations of $I_{pp}$ for
hadron-tagged jets accounting for the jet energy fluctuations
and the quenching effects for both the back-to-back jets.
In the present paper, we carry out such calculations of $I_{pp}$
for hadron-tagged
jets for conditions of the ALICE experiment \cite{ALICE_Ipp}
within the LCPI approach \cite{LCPI1} to the induced gluon emission. 
We calculate the medium modified FFs using the LCPI scheme
with temperature dependent $\alpha_s$ \cite{RAA20T}
used in our recent 
global analysis \cite{Z_hl}  of the data on the nuclear
modification factor $R_{AA}$.

{
  In the present analysis we ignore the possible
  effect on $I_{pp}$ of the double parton scattering (DPS)
with production the two nearly collinear
di-jets. The production of such di-jet pairs potentially
can lead to some dependence of $I_{pp}$ on the UE multiplicity due
to correlations between the UE activity and intensity of the DPS
\cite{Field}.
However, one can easily understand that the multiplicity dependence
of $I_{pp}$ observed in \cite{ALICE_Ipp} can not be due to the DPS
because its effect should be of the opposite sign.
Indeed, for the ALICE measurement we have $p_T^t/p_T^a\sim 2$,
therefore, due to the steep decrease of the jet cross sections
with the transverse momenta, the DPS contribution to $I_{pp}$
should be dominated by the production of two di-jets with different
energies, when the di-jet with the smallest energy contributes
mostly to the production of the associated hadrons. Since it is reasonable
to assume that the probability of the DPS should be bigger for events
with high UE multiplicity, we conclude that the DPS mechanism should
lead to a growth of $I_{pp}$ with $dN_{ch}/d\eta$, i.e. the effect
should be of opposite sign to that observed in \cite{ALICE_Ipp}.
Our estimates show that the magnitude of the DPS contribution to $I_{pp}$
should not exceed $\sim 1-2$\%. This conclusion is supported by
the recent simulations of $I_{pp}$ \cite{Ortiz} for
conditions of the ALICE measurement \cite{ALICE_Ipp}
within the PYTHIA Monte Carlo model, which includes the
DPS, where it was found that 
$I_{pp}$ in the away region is consistent with unity, and independent of
the UE multiplicity.
}\\

{\bf Theoretical framework for calculation of di-hadron and one-hadron
  cross sections}.
We calculate the di-hadron and one-hadron cross sections neglecting the
internal parton transverse momenta in the colliding protons.
In this case, in the approximation of independent
hadron fragmentation for the back-to-back jets, the cross sections
that we need can be written as
\bea
\frac{d^4\sigma_{m,v}}{dp_T^adp_T^tdy^ady^t}
=\int \frac{dz^t}{z^t}D_{h^t/i}^{m,v}(z^t,p_{Ti})\nonumber\\
\times D_{h^a/j}^{m,v}(z^a,p_{Tj})
\frac{d^3\sigma_{ij}}{p_{Ti}dp_{Ti}dy_idy_j}\,,
\label{eq:40}
\eea
\beq
\frac{d^2\sigma_{m,v}}{dp_T^tdy^t}
=\int \frac{dz^t}{z^t}D_{h^t/i}^{m,v}(z^t,p_{Ti})
\frac{d^2\sigma_{i}}{dp_{Ti}dy_i}\,,
\label{eq:50}
\eeq
where
$\frac{d^3\sigma_{ij}}{dp_{Ti}dy_idy_j}$
is the two-parton cross section of $p+p\to i+j+X$ process for
$y_i=y^t$, $y_j=y^a$,
$p_{Ti}=p_{Tj}=p_{T}^t/z^t$ and $z^{a}=z^{t}p_T^a/p_T^t$, 
$\frac{d^2\sigma_{i}}{dp_{Ti}dy_i}$
is the one-parton cross section for $p+p\to i+X$ process,
and $D_{h^t/i}$, $D_{h^a/j}$
are the $i\to h^t$ and $j\to h^a$ FFs.
The averaging over the jet trajectories
is implicit in (\ref{eq:40}) and (\ref{eq:50}). As in
our calculations of $R_{pp}$ in \cite{Z_hl},
we perform calculations of $I_{pp}$ for an effective azimuthally symmetric
mQGP fireball for the whole range of the impact parameter.
This approximation is reasonable because we calculate 
an azimuthally averaged quantity, and anyway
the jet production is dominated by $pp$ collisions
with small impact parameters.
As in \cite{Z_hl}, the distribution in the transverse plane of
the di-jet production points is evaluated using the MIT bag model quark density
(for simplicity we assume that gluons have the same distribution
in the transverse plane as quarks).

For calculation of the medium-modified FFs $D_{h/i}^{m}(z,Q)$
we use the same method as in \cite{Z_hl}. 
We write $D_{h/i}^{m}$ 
as the triple $z$-convolution 
\beq
D_{h/i}^{m}(Q)\approx D_{h/j}(Q_{0})
\otimes D_{j/k}^{in}\otimes D_{k/i}^{DGLAP}(Q)\,,
\label{eq:60}
\eeq
where $D_{k/i}^{DGLAP}$ is the DGLAP FF for $i\to k$ transition,
$D_{j/k}^{in}$ is the in-medium $j\to k$ FF,
and  $D_{h^{a,t}/j}$ describe vacuum transition of the parton $j$ to
hadrons $h^{a,t}$.
As in \cite{Z_hl}, we calculate the vacuum FFs $D_{h/i}^{v}(z,Q)$
dropping the induced stage FFs $D_{j/k}^{in}$ in (\ref{eq:60}).
To calculate the DGLAP FFs
$D_{k/i}^{DGLAP}$ we use the PYTHIA event generator \cite{PYTHIA}.
The in-medium FFs $D_{j/k}^{in}$ have been obtained in the approximation
of the independent gluon emission \cite{RAA_BDMS} using
the LCPI form of the induced gluon spectrum.
In calculation of $D_{j/k}^{in}$, we account for the
collisional energy loss, that is relatively small \cite{Z_coll,Gale_coll},
by treating it
as a perturbation to the radiative mechanism (see \cite{Z_hl}
for details).  

For the hadronization FFs $D_{h/j}$ we use the KKP
\cite{KKP} parametrization  with $Q_0=2$ GeV.
Note that if the trigger hadron $h^t$ is the leading
particle in the jet (as for the ALICE data \cite{ALICE_Ipp}),
then in (\ref{eq:60}) the ordinary FF
$D_{h^t/j}$
should be replaced by the probability distribution for production of the
leading hadron, $D_{h^t/j}^l=S_{h^t/j}D_{h^t/j}$.
For $z>0.5$ we have $S_{h^t/j}=1$, but at $z<0.5$
$S_{h^t/j}<1$ \cite{Renk1}. The behavior of $S_{h^t/j}$ at $z<0.5$
depends on the hadronization model.
Numerical simulation with PYTHIA event generator for the LUND fragmentation
gives $S_{h^t/j}$ that is close to unity at $z\gsim 0.3(0.2)$
for quarks(gluons),
and at smaller $z$ it falls steeply to zero.
We use a step-function parametrization
$S_{h^t/j}=\theta(z-z_0)$ with the cutoff parameter $z_0$ defined from
the normalization condition $\int_{z_0}^1 dz D_{h^t/j}(z)=1$ (such a
normalization is reasonable since the probability of jet hadronization
without charged particle production is $\ll 1$), that
leads to practically the same results for $I_{pp}$ as $S_{h^t/j}$
obtained using the PYTHIA event generator.
Note that, in principle, the presence of the suppression factor 
$S_{h^t/j}$ for the leading trigger hadron is not very important, since
it has a quite small effect on $I_{pp}$.
This occurs due to the steep fall-off of
the partonic cross sections, resulting in strong  suppression of the
small-$z$ contributions in formulas (\ref{eq:40})
and (\ref{eq:50}).

We calculate the gluon induced spectrum, which is needed
for calculation of $D_{j/k}^{in}$,
using for running $\alpha_s$ the same parametrization
as in \cite{RAA20T,Z_hl}
(supported by the lattice results for the in-medium $\alpha_s$ in the
coordinate space \cite{Bazavov_al1})
\beq
\alpha_s(Q,T) = \begin{cases}
\dfrac{4\pi}{9\log(\frac{Q^2}{\Lambda_{QCD}^2})}  & \mbox{if } Q > Q_{fr}(T)\;,\\
\alpha_{s}^{fr}(T) & \mbox{if }  Q_{fr}(T)\ge Q \ge cQ_{fr}(T)\;, \\
\frac{Q\alpha_{s}^{fr}(T)}{cQ_{fr}(T)} & \mbox{if }  Q < cQ_{fr}(T)\;, \\
\end{cases}
\label{eq:70}
\eeq
with $c=0.8$,
$Q_{fr}(T)=\Lambda_{QCD}\exp\left\lbrace
{2\pi}/{9\alpha_{s}^{fr}(T)}\right\rbrace$ (we take $\Lambda_{QCD}=200$ MeV)
and $Q_{fr}=\kappa T$, where $\kappa$ is a free parameter.
We take $\kappa=2.55$, which  has been obtained
by $\chi^2$ fitting of the LHC data on $R_{AA}$ for 2.76 and 5.02 TeV
Pb+Pb, and 5.44 TeV Xe+Xe collisions for scenario with the mQGP formation
in $pp$ collisions 
(in this scenario the theoretical $R_{AA}$ reads
$R_{AA}=R_{AA}^{st}/R_{pp}$,
where $R_{AA}^{st}$ is the standard  nuclear modification factor 
calculated with the pQCD $pp$ cross section, and $R_{pp}$ 
accounts for the medium jet modification
in the mQGP produced in $pp$ collisions, see \cite{Z_hl} for details).

{
  As in \cite{Z_hl}, we calculate $D_{j/k}^{in}$ for 
a fireball with a uniform entropy/density distribution in the transverse plane.
We checked that for $R_{pp}$ this approximation gives practically the
same results as calculations
with the gaussian entropy distribution (with the same mean square radius
and the total entropy). It is reasonable to expect that for $I_{pp}$
the accuracy of this approximation should be quite good as well.
This is connected with the fact 
that for $pp$ collisions jet quenching is dominated by the emission of
gluons with the formation length which is larger or of the order of the
mQGP fireball size.
In this regime the induced gluon emission is  mostly sensitive to
the total amount of the matter traversed by fast parton,
and the density profile along its trajectory is of minor importance.

It is worth noting that possible an incomplete equilibration/isotropization of
the matter in the initial stage \cite{RMK1512,Kurkela} is practically
irrelevant for the induced
gluon emission, because it is practically sensitive only to the parton number
density of the matter (see discussion in \cite{Z_pp13}). But of course, the
collectivity of the soft partons produced in $pp$ collision is crucial,
because it guarantees that they do not escape the interaction
region in the free-stream regime without interaction with the jets.
}\\

{\bf Model of the mQGP fireball}.
We assume the Bjorken 1+1D longitudinal expansion \cite{Bjorken}
of the QGP at the proper time $\tau>\tau_0=0.5$ fm, that
gives the entropy density $s=s_0(\tau_0/\tau)$, and for $\tau<\tau_0$
we take $s=s_0(\tau/\tau_0)$.
We calculate $I_{pp}$ for two versions of the initial entropy
density $s_0$. In the first version (A) 
we assume that the UE charged particles are produced from decay of a
thermalized mQGP fireball and write $s_{0}$ in terms of
the UE charged multiplicity density $dN_{ch}/d\eta$
as 
\beq
s_{0}=\frac{C}{\tau_{0}\pi R_{f}^{2}}\frac{dN_{ch}}{d\eta}\,,
\label{eq:80}
\eeq
where  $R_{f}$ is the fireball radius, and
$C=dS/dy{\Big/}dN_{ch}/d\eta\approx 7.67$ is the entropy/multiplicity
ratio \cite{BM-entropy}.  
As in \cite{Z_hl}, we calculate $R_f$ using the parametrization of the
fireball radius $R_{f}$ obtained in the CGC picture from \cite{glasma_pp}.
In the interval $dN_{ch}/d\eta\sim 5-20$, which is of interest here,
it gives a monotonic increase of $R_f$ from $\sim 1.2$ to $\sim 1.5$ fm.
For the minimum bias UE multiplicity
$dN_{ch}/d\eta\approx 12.5$ (it can be obtained
by interpolating the ATLAS 
data \cite{ATLAS_UE_Nch} for $\sqrt{s}=0.9$ and $7$ TeV
assuming that $dN_{ch}/d\eta\propto s^{\delta}$),
we obtain $R_f\approx 1.48$ fm. With this $R_f$ using (\ref{eq:80})
we obtain the average initial fireball temperature $T_0\approx 226 $
MeV for the ideal
gas model entropy density, and $T_0\approx 256$ MeV for the lattice
entropy density \cite{t-lat}.
In the second version (B), to account for 
the fact that at a sufficiently low UE multiplicity density
the assumption of a complete thermalization of the mQGP
fireball may become invalid, 
we multiply the right-hand side of (\ref{eq:80}) by
a multiplicity dependent coefficient $K_{th}(dN_{ch}/d\eta)$.
Considering the results
of \cite{ALICE_strange,Camp1}, that support
the onset of the QGP regime at $dN_{ch}/d\eta \sim 5-6$,
we use the parametrization
$K_{th}(N)=[1+\tanh((N-N_0)/D)]/2$ with $N_0=7.5$ and $D=2$. It
mimics the evolution of the jet quenching effect from very small
(at $dN_{ch}/d\eta \sim 5$)
to almost its ordinary magnitude for a thermalized mQGP fireball
(at $dN_{ch}/d\eta \gsim 10$).
It is worth noting that the results of our previous analysis \cite{Z_hl} of
the available
data on $R_{AA}$ from RHIC and the LHC also support the scenario
where in $pp$ collisions the mQGP is formed at $dN_{ch}/d\eta \gsim 10$,
but is absent for $dN_{ch}/d\eta \sim 5$,  because it is this very scenario that
allows to get the best simultaneous description of the data at the 
RHIC and the LHC energies (note that the minimum bias UE
$dN_{ch}/d\eta \sim 6$ and $\gsim 10$ in these cases). 

\begin{figure} [t]
\vspace{.7cm}
\begin{center}
\epsfig{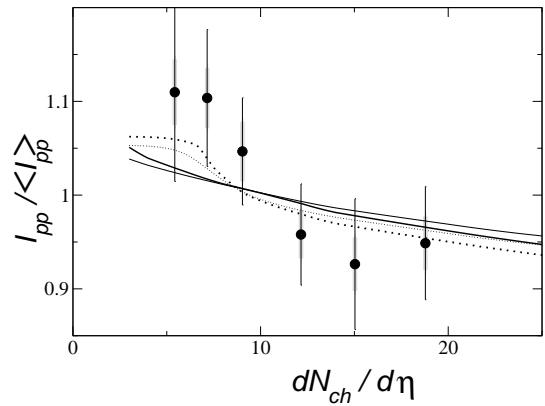}
\end{center}
\vspace{-0.5cm}
\caption[.]
{The ratio $I_{pp}/\langle I_{pp}\rangle$ vs the UE charged multiplicity
  density $dN_{ch}/d\eta$  for $pp$ collisions at $\sqrt{s}=5.02$ TeV.
  The solid and dotted curves are obtained
  in the versions A and B, respectively.
  Thin lines show the predictions obtained with the ordinary FF for
  the trigger particle, and the thick lines show the results obtained
  with the leading trigger particle probability distribution
(see text for explanation).  
Data points are from the preliminary ALICE measurements \cite{ALICE_Ipp}
(rescaled by the factor $r\approx 4.4$ to have the minimum bias
UE charged multiplicity density $dN_{ch}/d\eta \approx 12.5$
in the whole ranges of the transverse momentum and the azimuthal angle).
}
\end{figure}

{
  The approximation of the effective fireball, used in our jet quenching
  calculations,
  ignores fluctuations of the size and density of the mQGP,
  which fluctuate with the impact parameter of the $pp$ collisions and for
  each impact parameter. However, the effect of such fluctuations
  should be small. Indeed, for a small size QGP the energy loss
  grows approximately linearly with the fireball size and the QGP density
  \cite{Z_pp13}. The latter is due to dominance of the $N=1$ rescattering
  contribution for a small size QGP
  (see discussion of the pattern of induced gluon emission
  in a small size QGP in \cite{Z_pp13}). These facts  guarantee that
  the results for $I_{pp}$ should be weakly sensitive
  to the size and density fluctuations of the fireball, and justify
  the use the model of an effective fireball with fixed size and density.
}\\   

{\bf Numerical results}.
In Fig. 1 we plot the multiplicity dependence of the ratio $I_{pp}/\langle
I_{pp}\rangle$ (here, as in (\ref{eq:30}),
$\langle I_{pp}\rangle$ is the minimum bias value of $I_{pp}$)
obtained
for the versions A (solid) and B (dotted) of the initial entropy $s_0$.
To illustrate
the sensitivity to the trigger hadron fragmentation model, we show the
results obtained with the ordinary FF $D_{h^t/j}$ (thin lines) and with the
probability distribution of the leading
trigger particle $D_{h^t/j}^l$ (thick lines).
We evaluate $\langle I_{pp}\rangle$ using the multiplicity dependence
of the UE activity from CMS \cite{CMS_UE7} obtained for $7$ TeV $pp$
collisions (rescaled to $\sqrt{s}=5.02$ TeV assuming the KNO scaling).
As can be seen from
Fig. 1, our predictions are compatible within the errors with the ALICE data
\cite{ALICE_Ipp}. However, in both the versions the theoretical curves
have somewhat less steeper decrease with increasing multiplicity
than the data.
The discrepancy becomes smaller in the version B with an incomplete
thermalization at $dN_{ch}/d\eta \lsim 10$.
As one can see, the difference between 
the results  for our two trigger particle fragmentation models
(thin and thick lines) 
is relatively small.\\

{\bf Summary}.
We have investigated the quenching effects for
hadron-tagged jets in $pp$ collisions.
We have calculated the medium modification factor $I_{pp}$
within the LCPI approach to induced gluon emission
for parametrization of the running QCD coupling $\alpha_s(Q,T)$
which has a plateau around $Q\sim \kappa T$
(motivated by the lattice calculations of the effective
QCD coupling in the QGP \cite{Bazavov_al1}).
We use the value of $\kappa$ fitted to
the LHC data on the nuclear modification factor $R_{AA}$
in $2.76$ and $5.02$ TeV Pb+Pb, and $5.44$ TeV Xe+Xe collisions.
We find that the theoretical predictions with no free parameters
for
the multiplicity dependence of the ratio $I_{pp}/\langle I_{pp}\rangle$
for $5.02$ TeV $pp$ collisions
are in reasonable agreement with the recent preliminary data
from ALICE \cite{ALICE_Ipp}.
The description of the data becomes better
for the scenario with an incomplete
thermalization of the matter at $dN_{ch}/d\eta \lsim 10$.
Our results show that the drop of the ratio $I_{pp}/\langle I_{pp}\rangle$
with the UE multiplicity, if confirmed by further measurements,
may be viewed as the first direct evidence for the jet quenching
in $pp$ collisions.

\vspace{-.5cm}

\begin{acknowledgments}
  I am grateful to S.~Tripathy
  for useful communication on some aspects of the
ALICE measurement of  $I_{pp}$   
\cite{ALICE_Ipp}. 	
This work was supported by the Russian Science Foundation
under grant No~20-12-00200.
\end{acknowledgments}

\end{document}